\documentclass[aps, amsmath, amssymb, amsfonts]{article}
\usepackage{graphicx}
\usepackage{graphics}
\usepackage{dcolumn}
\usepackage{bm}
\usepackage{amssymb}
\usepackage{amsmath}
\usepackage{amsfonts}
\usepackage{amsopn}
\usepackage{epsfig}

\def\openone{\leavevmode\hbox{\small1\kern-0.8ex\normalsize1}}

\newcommand {\el} {\\ \nonumber}
\newcommand {\bra} [1] {\langle #1 |}
\newcommand {\ket} [1] {| #1 \rangle}
\newcommand {\bkt} [1] {\langle #1 \rangle}
\newcommand {\dbkt} [2] {\langle #1 | #2 \rangle}
\newcommand {\tbkt} [3] {\langle #1 | #2 | #3 \rangle}
\newcommand {\pd} [2] {\frac{\partial #1}{\partial #2}}
\newcommand {\td} [2] {\frac{d #1}{d #2}}

 \newcommand {\beq}{\begin{equation}}
\newcommand {\eeq}{\end{equation}}
\newcommand {\bea}{\begin{eqnarray}}
\newcommand {\eea}{\end{eqnarray}}
\begin{document}
\begin{center}
\title{Semiclassical spin transport in spin-orbit-coupled systems}
\maketitle
Dimitrie Culcer \footnote{Current affiliation: Condensed Matter Theory Center, Department of Physics, University of Maryland, College Park MD 20742.} \\
Advanced Photon Source, Argonne National Laboratory, Argonne, IL
60439 \\ Northern Illinois University, De Kalb, IL 60115
\end{center}
\begin{abstract}
This article discusses spin transport in systems with spin-orbit
interactions and how it can be understood in a semiclassical
picture. I will first present a semiclassical wave-packet
description of spin transport, which explains how the microscopic
motion of carriers gives rise to a spin current. Due to spin
non-conservation the definition of the spin current has some
arbitrariness. In the second part I will briefly review the physics
from a density matrix point of view, which makes clear the
relationship between spin transport and spin precession and the
important role of scattering.
\end{abstract}

\tableofcontents

\section*{Glossary}

\begin{itemize}

\item \textit{Extrinsic effect} an effect which has an explicit dependence on the form or strength of the
disorder potential.

\item \textit{Intrinsic effect} an effect which does not depend explicitly on the form and strength of the
disorder potential.

\item \textit{Semiclassical theory} a theory in which a particle's position and
momentum are considered simultaneously.

\item \textit{Spin-orbit interaction} a relativistic interaction
between the spin of a particle and its momentum (which is associated
with its orbital motion.)

\item \textit{Steady-state spin current} a flow of spins induced by
an electric field.

\item \textit{Steady-state spin density} a net spin density induced by
an electric field.

\end{itemize}

\section{Definition of the subject and its importance}

\emph{Spin transport} refers to the physical movement of spins
across a sample and, if spin were a conserved quantity, one could
make a straightforward distinction between spin-up and spin-down
charge currents. The recent upsurge of interest in spin transport
is, however, motivated by systems in which spin is not conserved due
to the presence of spin-orbit interactions, which give rise to spin
precession. Here, due to non-conservation of spin the spin current
is not well defined \cite{shi06, sug06, wang06}. Spin transport in
these cases usually does not involve charge transport as the charge
currents in the direction of spin flow cancel out. Finally, in
certain materials, spin currents are accompanied by steady-state
spin densities. The appearance of a spin density is not a transport
phenomenon, but it is a steady-state process and is intimately
connected to spin transport.

The word \emph{semiclassical} as used in this work refers to
theories which consider the position and momentum of a particle
simultaneously. Semiclassical pictures are intuitive and useful in
descriptions of transport, particularly in inhomogeneous systems and
in spatially dependent fields, which typically vary on length scales
much larger than atomic size.

In recent years, steady progress has been made towards realization
of convenient semiconducting ferromagnets and spin injection into
semiconductors from ferromagnetic metals \cite{Fiederling, OhnoII,
Parkin, Jonker} yet spin injection from a ferromagnetic metal into a
semiconductor is hampered by the resistivity mismatch between the
two \cite{Schmidt}. This is one factor, in addition to basic
science, motivating the search for an understanding of the way spins
are manipulated electrically. The last few years have seen many
experimental advances in spin transport, and spin currents have been
measured directly \cite{val06, liu06} and indirectly \cite{kato04,
wun05, sih05, ste06, gan01}.

\section{Introduction}

Novel physical phenomena that may lead to improved memory devices
and advances in quantum information processing are closely related
to spin-orbit interactions. \cite{zut04} Spin-orbit interactions are
present in the band structure and in potentials due to impurity
distributions. Spin-orbit coupling is in principle always present in
impurity potentials and gives rise to skew scattering. Band
structure spin-orbit coupling may arise from the inversion asymmetry
of the underlying crystal lattice \cite{dre55} (bulk inversion
asymmetry), from the inversion asymmetry of the confining potential
in two dimensions \cite{ras84} (structure inversion asymmetry), and
may be present also in inversion symmetric systems. \cite{lut56}

Although many observations in this entry are general, the discussion
will focus on non-interacting spin-1/2 electron systems, which are
pedagogically easier. The Hamiltonian of these systems typically
contains a kinetic energy term and a spin-orbit coupling term,
$H_{{\bm k}} = \frac{\hbar^2k^2}{2m^*} + H_{{\bm k}}^\mathrm{so}$,
where $m^*$ is the electron effective mass. In spin-1/2 electron
systems, band structure spin-orbit coupling can always be
represented as a Zeeman-like interaction of the spin with a wave
vector-dependent effective magnetic field ${\bm \Omega}_{\bm k}$,
thus $H_{{\bm k}}^\mathrm{so} = (1/2) \,{\bm \sigma}\cdot
{\bm\Omega}_{\bm k}$. Common examples of effective fields are the
Rashba spin-orbit interaction, \cite{ras84} which is often dominant
in quantum wells with inversion asymmetry, and the Dresselhaus
spin-orbit interaction, \cite{dre55} which is due to bulk inversion
asymmetry. The spin operator is given by $s^\sigma = (\hbar/2)\,
\sigma^{\sigma}$, where $\sigma^\sigma$ is a Pauli spin matrix. The
spin current operator in these systems will be taken to be
$\hat{\mathcal{J}}^\sigma_{i} = (1/2)\, \{ s^\sigma, v^i \}$, where
the velocity operator is $v^i = (1/\hbar) \, \partial H_{\bm
k}/\partial k_i$.

An electron spin at wave vector ${\bm k}$ precesses about the
effective field ${\bm \Omega}_{\bm k}$ with frequency
$\Omega_{\bm{k}}/\hbar \equiv |{\bm \Omega_{\bm{k}}}|/\hbar$ and is
scattered to a different wave vector within a characteristic
momentum scattering time $\tau_p$. I will assume in this work that
$\varepsilon_F\tau_p/\hbar \gg 1$, where $\varepsilon_F$ is the
Fermi energy, which is equivalent to the assumption that the carrier
mean free path is much larger that the de Broglie wavelength. Within
this range, the relative magnitude of the spin precession frequency
$\Omega_{\bm{k}}$ and inverse scattering time $1/\tau_p$ define
three qualitatively different regimes. In the ballistic (clean)
regime no scattering occurs and the temperature tends to absolute
zero, so that $\varepsilon_F\tau_p \rightarrow \infty$ and
$\Omega_{\bm{k}}\tau_p/\hbar \rightarrow \infty$. The weak
scattering regime is characterized by fast spin precession and
little momentum scattering due to, e.g., a slight increase in
temperature, yielding $\varepsilon_F\tau_p/\hbar \gg \Omega_{\bm{k}}
\, \tau_p/\hbar \gg 1$. In the strong momentum scattering regime
$\varepsilon_F\tau_p/\hbar \gg 1 \gg \Omega_{\bm{k}} \, \tau_p/\hbar
$. I will concentrate on effects originating in the band structure,
the observation of which requires the assumption that the materials
under study are in the weak momentum scattering regime. Electric
fields will be assumed uniform.

The first part of this article will present a semiclassical theory
of spin transport, identifying the terms responsible for spin
currents in the microscopic dynamics of carriers. Spin
non-conservation as a result of spin precession leads to several
possible definitions of the spin current, which emerge out of the
spin equation of continuity. The second part presents a different
point of view, which explains aspects not easily captured in the
semiclassical approach. The steady-state density matrix is shown to
contain a contribution due to precessing spins and one due to
conserved spins. Steady state corrections $\propto \tau_p$ are
associated with the \emph{absence} of spin precession and give rise
to spin densities in external fields. \cite{ivc78, lev85, ede90,
aro91, cha01, kato04b, vor79, gan04, sil04} Steady state corrections
independent of $\tau_p$ are associated with spin precession and give
rise to spin currents in external fields. \cite{shi06, sug06,
wang06, val06, liu06, kato04, wun05, sih05, ste06, gan01, dya71,
hir99, Zhang, QZh, eng05, mur03, sin04, ada05, ras04, zha05, ino04,
dim05, sch02, mis04, kha06, shy06, ll06, mal05} Scattering between
these two distributions induces significant corrections to
steady-state spin currents.

\section{Spin currents in electric fields}

\subsection{Wave-packet picture of spin transport}

This section presents a semiclassical theory of spin transport valid
for a general spin-orbit system. The semiclassical method is a
suitable approach to the study of transport, because, typically, in
the relevant systems the external fields vary smoothly on atomic
length scales. All information about the system is taken to be
contained in the band structure, thus allowing a description of spin
transport which does not make reference to the detailed form of the
spin-orbit interaction.

The system under study is regarded as as a collection of carriers,
whose semiclassical dynamics in a non-degenerate band $i$ are
described by a wave packet \cite{Sundaram}, with its charge centroid
having coordinates (${\bf r}_c, {\bf k}_c$) \beq |w_i \rangle=\int
d^3k \; a({\bf k}, t) e^{i{\bf k}\cdot\hat{\bf r}}|u_i({\bf r}_c,
{\bf k}, t)\rangle . \eeq In the above, the function $a({\bf k}, t)$
is a narrow distribution sharply peaked at ${\bf k}_c$, the phase of
which specifies the center of charge position ${\bf r}_c$, while
$|u_i({\bf r}_c, {\bf k}, t)\rangle$ are lattice-periodic Bloch wave
functions. The size of the wave packet in momentum space must be
considerably smaller than that of the Brillouin zone. In real space,
this implies that the wave packet must stretch over many unit cells.

The external electric field drives the center of the wave packet in
$k$-space according to the semiclassical equations of motion \bea
\hbar \dot {\bf r}_c = \pd{\varepsilon_i}{{\bf k}_c} - q{\bf E}
\times {\bf \Omega}_i \el \hbar \dot {\bf k}_c = q{\bf E}
\label{sc}, \eea with $q$ the charge of the carriers,
$\varepsilon_i$ the band energy, and the Berry curvature \beq {\bf
\Omega}_i = i \bra{\pd{u_i}{{\bf k}}}\times\ket{\pd{u_i}{{\bf k}}}.
\eeq The electric field also gives rise to an adiabatic correction
to the wave functions, which mixes the states making up the wave
packet. The wave functions $\ket{u_i}$ therefore have the following
form: \beq \label{uptb} \ket{u_i} = \ket{\phi_i} -
\sum_j\frac{\tbkt{\phi_j}{i\hbar\td{}{t}}{\phi_i}}{\varepsilon_i -
\varepsilon_j}\ket{\phi_j}, \eeq where the $\phi_i$ are the
unperturbed Bloch eigenstates. The $\ket {u_i}$ form a complete set
and retain the Bloch periodicity.

The distribution of carriers is described by a function $f$. When
scattering is present, the distribution function satisfies the
following equation: \beq \pd{f}{t} + \dot{\bf r}_c\cdot\pd{f}{{\bf
r}_c} + \dot{\bf k}_c\cdot\pd{f}{{\bf k}_c} = \bigg(\td{f}
{t}\bigg)_{coll}, \label{Boltz} \eeq where $(\td{f} {t})_{coll}$ is
the usual collision term. In independent bands, in the relaxation
time approximation, the collision term takes the form $\frac{f_0 -
f}{\tau_p}$, with $f_0$ the equilibrium distribution and $\tau_p$
the momentum relaxation time. In the Boltzmann theory, the change in
the distribution function with time arises through the drift terms,
which are determined from the semiclassical equations of motion, as
well as through scattering with other carriers, with localized
impurities or with phonons. For transport in a non-degenerate band,
it is consistent to ignore interband scattering effects in the weak
scattering limit. In this case the relaxation time is a scalar
quantity. The effects of interband coherence due to scattering will
be explored in the next section.

In order to obtain expressions for macroscopic quantities of
interest, such as densities and currents, one needs to carry out a
coarse graining by averaging over microscopic fluctuations. In
classical dynamics this coarse graining is performed by means of a
sampling function, which is smooth and has a significant magnitude
only in a finite range \cite{Jackson}. This range is large compared
to atomic dimensions, but small compared to the scale of variation
of the distribution function. Moreover, it has a rapidly converging
Taylor expansion over distances of atomic dimensions, and its form
does not need to be specified. This method has a close analog in
wavepacket dynamics, where the sampling function is replaced by a
$\delta$-function.

\begin{figure}[tbp]
\centering \epsfig{file=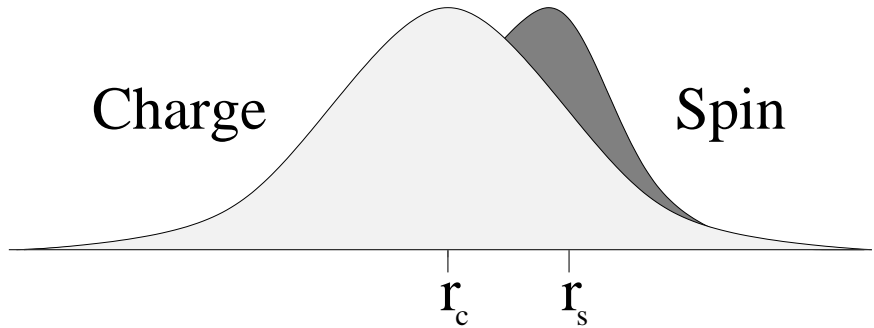, width=0.95\columnwidth}
\caption{For a particle of finite extent the charge and spin
distributions in real space are in general do not coincide. The same
is true of the charge and spin distributions in reciprocal space.}
\label{fig:wavepacket}
\end{figure}

It is crucial to recognize that, in general, the center of spin and
the center of charge are distinct, since the wave packet samples a
range of wave vectors and the spin is usually a function of {\bf k}.
Following the line of thought outlined above, the spin density is
defined to be (henceforth ${\bf k}_c$ will be abbreviated to {\bf
k}) \beq S^\sigma({\bf R}, t) = \int\!\int d^3k d^3r_c f({\bf r}_c,
{\bf k}, t)\bkt{\delta({\bf R} - {\bf \hat r})\hat {s}^\sigma}, \eeq
where the bracket indicates quantum mechanical averaging over the
wave packet with charge centroid $({\bf r}_c, {\bf k})$. As the
$\delta$-function has operator arguments, it will be regarded as a
\emph{sampling operator}, whose expectation value yields a spatial
average, evaluated at position {\bf r}. To account for the fact that
spin is not conserved, a new quantity is introduced, which will be
referred to as the torque density, defined by \beq {\cal T}^\sigma
({\bf R}, t) = \int\!\int d^3k d^3r_c f({\bf r}_c, {\bf k},
t)\bkt{\delta({\bf R} - {\bf \hat r})\hat{\tau}^\sigma}. \eeq
$\hat{\tau}^\sigma$ in the above stands for the rate of change of
the spin operator, given by $\frac{i}{\hbar}[\hat H, \hat
{s}^\sigma]$, and symmetrization of products of non-commuting
operators has been assumed. Finally, the microscopic spin current
density is defined as: \beq \bm{\mathcal{J}}^{\sigma}({\bf R}, t) =
\int \!\int d^3k d^3r_c f({\bf r}_c, {\bf k}, t)\bkt{\delta({\bf R}
- {\hat{\bf r}}){ \hat{s}^\sigma{\hat{\bf v}}} } . \eeq We obtain
the following continuity equation for the spin density and current:
\beq \label{continuity} \pd{S^\sigma}{t} + \nabla\cdot\bm{\mathcal
J}^\sigma = {\cal T}^\sigma + {\cal F}^\sigma. \eeq The equation of
continuity contains a bulk source term, which coincides with the
torque density and acts as a mechanism for spin generation. Similar
source terms are associated with nonconserved quantities, for
example, in quantum electrodynamics and in Maxwell's equations. The
last term in (\ref{continuity}) represents the scattering
contribution, which will be discussed further below.

\begin{figure}[tbp]
\centering \epsfig{file=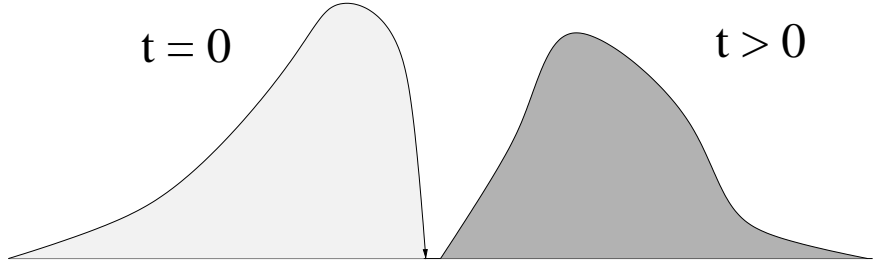, width=0.95\columnwidth}
\caption{In the presence of spin-orbit interactions the spin
distribution of a particle changes in time. The horizontal axis may
represent position or wave vector.} \label{fig:spinintime}
\end{figure}

Let us discuss the terms in the equation of continuity, beginning
with the spin density. The argument of the sampling operator can be
expressed as $[{\bf r} - {\bf r}_c - (\hat{\bf r} - {\bf r}_c)]$,
and, as the second term is of atomic dimensions, the sampling
operator can be written as a Taylor expansion about $(\hat {\bf r} -
{\bf r}_c)$. The density can therefore be re-expressed, in terms of
macroscopic quantities, as \beq S^\sigma({\bf R}, t) =
\rho^{s\sigma}({\bf R}, t) - \nabla_{\bf R}\cdot{\bf
P}^{s\sigma}({\bf R}, t) , \eeq where summation over repeated
indices has been assumed. In the above, the monopole density is
given by \bea \rho^{s\sigma}({\bf R}, t) = \int\!\int d^3k d^3r_c
f({\bf r}_c, {\bf k}, t)\bkt{\hat {s}^\sigma}\delta({\bf R} - {\bf
r}_c) = \int d^3k f\bkt{\hat s^\sigma}|_{{\bf r}_c = {\bf R}}, \eea
where $f$ in the second line, and henceforth, is to be understood as
$f({\bf R}, {\bf k}, t)$, and the dipole density is \bea {\bf
P}^{s\sigma}({\bf R}, t) = \int\!\int d^3k d^3r_c f\bkt{(\hat{\bf r}
- {\bf r}_c){\hat {s}^\sigma}}\delta({\bf R} - {\bf r}_c) = \int
d^3k f{\bf p}^{s\sigma}|_{{\bf r}_c = {\bf R}}. \eea The average
spin of the wave packet has been denoted by $\bkt{\hat {s}^\sigma}$,
and the spin-dipole is defined to be ${\bf p}^{s\sigma} =
\bkt{(\hat{\bf r} - {\bf r}_c)\hat {s}^\sigma}|_{{\bf r}_c = {\bf
R}}$. It will be seen that the first term in the density is the
average of a monopole density located at ${\bf r}_{c}$, while the
dipole term is the average of a point dipole density located at
${\bf r}_{c}$, and similarly for higher orders. The dipole must be
understood as the average of the quantum mechanical dipole operator,
as an exact analogy with the electric dipole of classical
electrodynamics cannot be made. The density can thus be viewed as a
collection of point multipoles, located at the centroid of each wave
packet. The microscopic distribution of spin is important at the
molecular level, but at the macroscopic level the effect of this
molecular distribution is replaced by a sum of multipoles. Since the
center of spin is different from the center of charge, in principle
all multipoles are present.

Following a similar manipulation and using the Boltzmann equation,
the torque density is re-expressed as:
\begin{equation}
{\cal T}^\sigma({\bf R}, t) = \rho^{\tau\sigma}({\bf R}, t) -
\nabla\cdot{\bf P}^{\tau\sigma}({\bf R}, t)
\end{equation}
with the torque monopole density \bea \rho^{\tau\sigma}({\bf R}, t)
= \int\!\int d^3k d^3r_c f({\bf r}_c, {\bf k},
t)\bkt{\hat{\tau}^\sigma}\delta({\bf R} - {\bf r}_c) = \int d^3k
f\bkt{\hat{\tau}^\sigma}|_{{\bf r}_c = {\bf R}}, \eea and the torque
dipole density \bea {\bf P}^{\tau\sigma}({\bf R}, t) = \int\!\int
d^3k d^3r_c f({\bf r}_c, {\bf k}, t)\bkt{({\bf \hat R} - {\bf r}_c)
\hat{\tau}^\sigma}|_{{\bf r}_c = {\bf R}} = \int d^3k f {\bf
p}^{\tau\sigma}|_{{\bf r}_c = {\bf R}}. \eea In analogy with the
spin dipole, the torque dipole has been defined as ${\bf
p}^{\tau\sigma} = \bkt{(\hat{\bf r} - {\bf r}_c)\hat
\tau^\sigma}|_{{\bf r}_c = {\bf R}}$. The torque density is
therefore also a sum of multipole moments, that is, the moments of a
point spin source located at ${\bf r}_c$. Even in the case when the
center of $\bkt{\hat {s}^\sigma}$ coincides with the center of
charge, $\bkt{\hat{\tau}^\sigma}$ may not be centered at ${\bf
r}_c$, with the result that the higher order terms in the torque
density are in general present. The second and higher terms of
${\cal T}^\sigma$ cancel exactly the analogous terms in the
continuity equation which come from the current.

Since only the gradient of the spin current appears in the equation
of continuity, in the expansion of the sampling operator we keep the
leading term \bea \label{js} \bm{\mathcal{J}}^s({\bf R}, t) = \int
d^3k f\bkt{{\hat{\bf v}}\hat {s}^\sigma}|_{{\bf r}_c = {\bf R}}.
\eea Keeping terms to first order in $({\bf \hat r} - {\bf r}_c)$,
the current can be decomposed into the following: \bea
\bm{\mathcal{J}}^{s\sigma} = {\bf C}^{s\sigma} + {\bf D}^{s\sigma} -
{\bf P}^{\tau\sigma} \eea The convective term ${\bf C}^{s\sigma}$
represents the spin being transported along with the wave packet
\bea {\bf C}^{s\sigma}({\bf R}, t) = \int\!\int d^3k d^3r_c f({\bf
r}_c, {\bf k}, t){\bf \dot r}_c \bkt{\hat {s}^\sigma}\delta({\bf R}
- {\bf r}_c) = \int d^3k f {\bf c}^{s\sigma}|_{{\bf r}_c = {\bf R}},
\eea while ${\bf D}^{s\sigma}$ comes from the rate of change of the
spin dipole, which has already been introduced. It has the form:
\bea {\bf D}^{s\sigma}({\bf R}, t) = \int d^3k f\td{{\bf
p}^{s\sigma}}{t}|_{{\bf r}_c = {\bf R}} \eea ${\bf P}^{\tau\sigma}$
is the torque dipole introduced above. The corresponding monopole
term appears in the source term of the continuity equation, as will
emerge below. The presence of the torque dipole here is to be
contrasted with the absence of an analogous term in classical
electrodynamics.  There, an electric dipole arises from the
placement of two charges a small distance from each other, but the
charge itself is conserved.

Finally, we come to the source term in (\ref{continuity}). The first
part, composed of the torque density, has already been discussed.
The second term, denoted by ${\cal F}^\sigma$, becomes, in the
relaxation time approximation \beq \label{reltime} {\cal F}^\sigma =
\frac{S^\sigma_0 - S^\sigma}{\tau_p} = \frac{1}{\tau_p}\int d^3k
(f_0 - f)\bkt{\hat {s}^\sigma}, \eeq where $\tau_p$ is the momentum
relaxation time and $f_0$ the equilibrium distribution, which is
usually the Fermi-Dirac distribution function.

Based on the continuity equation alone, there is some flexibility in
defining the current and the source. In systems in which spin is
conserved, the torque density becomes, to first order in $({\bf \hat
r} - {\bf r}_c)$, a pure divergence, which can be incorporated into
a redefinition of the spin current. This current, henceforth
referred to as the spin transport current, is only due to the
convective and spin dipole contributions: \beq {\bf
J}^{t\sigma}({\bf R}, t) = {\bf C}^{s\sigma}({\bf R}, t) + {\bf
D}^{s\sigma}({\bf R}, t) \label{jst} \eeq With respect to this spin
transport current, the continuity equation takes the following form:
\beq \label{stsource} \pd{S^\sigma}{t} + \nabla\cdot{\bf
J}^{t\sigma} = \td{}{t}\int d^3kf\bkt{\hat {s}^\sigma} \eeq

In the steady state under a constant electric field, the
distribution function is composed of an equilibrium part,
independent of the field, and a non-equilibrium part, which is first
order in the field. Henceforth, terms in the spin current and source
which depend on the equilibrium distribution function will be
referred to as intrinsic, whereas the terms depending on the
nonequilibrium shift in the distribution will be referred to as
extrinsic. For example, the integrand in Eq. (\ref{js}) can be
decomposed into a zero order spin-velocity, ${\bf v}\bkt{\hat
{s}^\sigma}$, where ${\bf v}$ is the usual group velocity of the
band, and a first order correction. Therefore, there will be a
contribution to the current from the non-equilibrium part of the
distribution and the zero order spin-velocity, which has been
discussed extensively in previous work \cite{dya71, hir99, Zhang,
QZh, eng05}. There will also be a contribution from the equilibrium
distribution and the first order correction to the spin-velocity,
which is referred to as the intrinsic contribution. In the wave
packet formalism presented here this effect arises from the change
in wave functions induced by the electric field, rather than from
the change in distribution functions that is responsible for most
conventional transport effects. The intrinsic spin current is
calculated from (\ref{js}) using the equilibrium distribution and
the expectation values of the spin and spin dipole operators in a
Bloch state perturbed to first order in ${\bm E}$.

In its turn, the source in (\ref{continuity}) can be decomposed into
intrinsic and extrinsic contributions. The present entry considers
homogeneous systems, so that all the gradient terms vanish, and the
torque density is simply $f\bkt{\hat {\tau}^\sigma}$. The zeroth
order contribution to this term is null, as the Bloch wave functions
are eigenstates of the Hamiltonian. Thus, to first order in the
electric field, we find that $\bkt{\hat {\tau}^\sigma}$ is simply
given by $(e{\bf E}/ \hbar)\cdot(\partial\bkt{\hat
{s}^\sigma}/\partial{\bf k})$. One is thus justified in replacing
$f$ by its equilibrium value $f_0$, in which case this term is
purely intrinsic. The second term in the source, ${\cal F}^\sigma$,
which depends on the nonequilibrium shift in the distribution
function, is entirely extrinsic.

The extrinsic source term takes into account the effect of
scattering, and is a term which usually appears in the equation of
continuity. The intrinsic source accounts for the effect of all
spin-nonconserving terms, and must be present even in a clean
system, if the Hamiltonian contains spin-dependent contributions. In
general, in addition to the rate of change of spin arising from the
spin-dependent terms in the Hamiltonian, scattering processes may
alter the orientation of the spin, with the result that any one spin
component is not conserved, and the orientation of spins is
randomized over a longer time period. For a uniform steady-state
system, the current is constant and the intrinsic source term must
vanish. However, near the boundary of the system, or at an interface
with a different semiconductor with (for example) weaker spin-orbit
interactions, the spin current driven by an electric field will vary
spatially and ${\cal T}$ must reach a non-zero value.

Let us take a closer look at the spin dipole and torque dipole,
which are seen to be the main mechanisms responsible for generating
the spin current. Because of its narrow distribution in ${\bf k}$,
the mean spin of the wave packet is $\langle w_i |\hat {s}^\sigma
|w_i \rangle =\langle u_i |\hat {s}^\sigma | u_i \rangle$, where it
is understood that the wave vector of the Bloch function is set at
${\bf k}_c$ and $\hat {s}^\sigma$ is an arbitrary projection of the
spin vector operator. The spin dipole of the wave packet, defined
relative to the charge center of the wave packet is given, in terms
of Bloch functions, by the expression:
\begin{equation}
 {\bf p}^{s\sigma}_i = \frac{i}{2}[\tbkt{u_i}{\hat{s}^\sigma}{\pd{u_i}{{\bf k}}} -
 \tbkt{\pd{u_i}{{\bf k}}}{\hat {s}^\sigma}{u_i}] - \dbkt{u_i}{i\pd{u_i}{{\bf k}}}\tbkt{u_i}{\hat {s}^\sigma}{u_i}
\label{cov}
\end{equation}
Interestingly, the spin dipole is independent of the wave packet
width. The expression is also invariant under a local gauge
transformation, in the sense that if $\ket{u_i}$ is modified by a
phase factor $e^{i\alpha({\bf k})}$ the spin dipole is unchanged.

The torque dipole term has a special interpretation in the case of
spin transport. The rate of change of spin is equivalent to a
torque, and the torque dipole represents the moment exerted by this
torque about the center of the wave packet. The semiclassical
expression for the torque moment is
\begin{equation}
{\bf p}^{\tau\sigma}_i  = \frac{i}{2}[\tbkt{u_i}{\dot {\hat
{s}}^\sigma}{\pd{u_i}{{\bf k}}} - \tbkt{\pd{u_i}{{\bf k}}}{\dot
{\hat {s}}^\sigma}{u_i}] - \dbkt{u_i}{i\pd{u_i}{{\bf
k}}}\tbkt{u_i}{\dot {\hat {s}}^\sigma}{u_i} \label{ts}.
\end{equation}
The torque moment has the same gauge invariance properties as the
spin dipole, and like the spin dipole it also does not depend on the
wave packet width.

It is important to note that the spin current,
$\bm{\mathcal{J}}^\sigma$, can be simplified to: \beq
\bm{\mathcal{J}}^\sigma({\bf R}, t) = \int d^3k \, f\, {\rm tr}\,
\tbkt{u_i}{\hat {s}^\sigma {\hat {\bf v}}}{u_i}, \eeq which is the
semiclassical equivalent of the Kubo formula for spin currents.

\subsection{Density matrix picture of spin transport}

Semiclassical theory provides a straightforward, intuitive picture
of the way spin currents arise in the course of carrier dynamics in
an electric field. The theory was developed for independent bands.
It turns out that interband coherence arising from scattering is
crucial in spin transport, and is difficult to treat
semiclassically. Although the semiclassical theory can be
generalized to multiple bands \cite{Coupled, wig06}, it is more
instructive to examine spin transport from a different point of view
that is closer in outlook to the philosophy underlying the Kubo
formula (with which the semiclassical theory agrees.) This will shed
some light on additional issues, such as the relationships between
spin currents and spin precession, between spin currents and spin
densities, the complex effect of disorder and the vanishing of spin
current in certain systems.

A large, uniform system of non-interacting spin-1/2 electrons is
represented by a one-particle density operator $\hat \rho$. The
expectation value of an observable represented by a Hermitian
operator $\hat O$ is given by ${\rm tr} (\hat\rho\hat O)$ and $\hat
\rho$ satisfies the quantum Liouville equation
\begin{equation}
\td{\hat\rho}{t} + \frac{i}{\hbar} \, [\hat H + \hat U, \hat \rho] =
0.
\end{equation}
The Liouville equation is projected onto a set of time-independent
states of definite wave vector $\{ \ket{{\bm k}s} \}$, which are not
assumed to be eigenstates of the Hamiltonian $\hat H$. The matrix
elements of $\hat \rho$ in this basis will be written as $\rho_{{\bm
k}{\bm k}'} \equiv \rho^{ss'}_{{\bm k}{\bm k}'} = \bra{{\bm k}s}
\hat\rho \ket{{\bm k}'s'}$. Spin indices are not shown explicitly,
and $\rho_{{\bm k}{\bm k}'}$ is a matrix in spin space, referred to
as the density matrix. In this work we require the expectation
values of operators which are diagonal in wave vector, and will thus
require the part of the density matrix diagonal in wave vector,
$\rho_{{\bm k}{\bm k}} \equiv f_{\bm k} = n_{\bm k} \openone +
S_{\bm k}$. In the presence of a constant uniform electric field
${\bm E}$, $f_{\bm{k}} = f_{0 \bm{k}} + f_{E {\bm{k}}} $, where the
equilibrium density matrix $f_{0 \bm{k}}$ is given by the
Fermi-Dirac distribution, and the correction $f_{E {\bm{k}}}$ is due
to the ${\bm E}$. We subdivide $f_{0{\bm k}} = n_{0{\bm k}} \,
\openone + S_{0{\bm k}}$ and $f_{E {\bm{k}}} = n_{E {\bm{k}}} \,
\openone + S_{E {\bm{k}}}$. The spin-dependent part of the
nonequilibrium correction to the density matrix $S_{E{\bm k}}$ is
interpreted as the spin density induced by ${\bm E}$. The equations
governing the time evolution of $n_{E{\bm k}}$ and $S_{E{\bm k}}$ is
\cite{dim07ss}
\begin{equation}\label{eq:SBoltz}
\begin{array}{rl}
\displaystyle \pd{n_{E{\bm k}}}{t} + \hat J_0 \, (n_{E {\bm{k}}}) =
& \displaystyle \frac{e{\bm E}}{\hbar}\cdot\pd{n_{0{\bm k}}}{{\bm
k}} \\ [3ex] \displaystyle \pd{S_{E {\bm{k}}}}{t} +
\frac{i}{\hbar}\, [H_{\bm k}, S_{E {\bm{k}}}] + \hat J_0 \, (S_{E
{\bm{k}}}) = & \displaystyle \frac{e{\bm E}}{\hbar}\cdot\pd{S_{0{\bm
k}}}{{\bm k}} - \hat J_s \, (n_{E {\bm{k}}}) \equiv \Sigma_{E
{\bm{k}}},
\end{array}
\end{equation}
where the scalar part of the scattering operator $\hat{J}_0$ and its
spin-dependent part $\hat{J}_s$ have been defined in Ref.\
\cite{dim07ss}. The equation for $n_{E{\bm k}}$ has the well-known
solution $n_{E{\bm k}} = (e{\bm E}\tau_p/\hbar)\cdot (\partial
n_{0{\bm k}}/\partial {\bm k})$, in other words, $n_{E{\bm k}}$
describes the shift of the Fermi sphere in the presence of the
electric field ${\bm E}$, with the momentum relaxation time
$\tau_p$. It is seen from Eq.\ (\ref{eq:SBoltz}) that spin-dependent
scattering gives rise to a renormalization of the driving term in
the equation for $S_{E{\bm k}}$. This renormalization has no analog
in charge transport.

We need to find the expectation value of the spin current operator
$\hat{\mathcal{J}}^\sigma_{i}$ defined in the introduction. In the
systems under study the spin current operator can be written as
$\hat{\mathcal{J}}^\sigma_{i} = \hbar k_i s^\sigma/m^* + (1/4\hbar)
\, \partial \Omega^\sigma/\partial k_i \openone$. We need to
determine $S_{E{\bm k}}$. To this end we remember that an electron
spin at wave vector ${\bm k}$ precesses about an effective magnetic
field ${\bm \Omega}_{\bm k}$. The spin can be resolved into
components parallel and perpendicular to ${\bm \Omega}_{\bm k}$. In
the course of spin precession the component of the spin parallel to
${\bm \Omega}_{\bm k}$ is conserved, while the perpendicular
component is continually changing. Corresponding to this
decomposition of the spin is an analogous decomposition of the spin
distribution $S_{E {\bm{k}}}$ into a part representing conserved
spin and a part representing precessing spin, denoted by $S_{E
{\bm{k}}\|}$ and $S_{E {\bm{k}}\perp}$ respectively. There is an
analogous decomposition of the source on the RHS of Eq.\
(\ref{eq:SBoltz}) into $\Sigma_{E {\bm{k}}\|}$ and $\Sigma_{E
{\bm{k}}\perp}$. This decomposition is carried out by introducing
projection operators $P_\|$ and $P_\perp$ as described in Ref.\
\cite{dim07ss}, giving for $S_{E {\bm{k}}\|}$ and $S_{E
{\bm{k}}\perp}$ in the weak momentum scattering limit
\begin{subequations}\label{subeq:S}
\begin{eqnarray}
& & \pd{S_{E {\bm{k}}\|}}{t}  +  P_\| \hat{J}_0 \, (S_{E {\bm{k}}})
= \Sigma_{E{\bm k}\|}, \label{eq:Sigmaparallel}
\\ [1ex]
& & \pd{S_{E {\bm{k}}\perp}}{t}  + \frac{i}{\hbar}\, [H_{\bm k},
S_{E {\bm{k}}\perp}] = \Sigma_{E{\bm k}\perp} - P_\perp \hat{J}_0 \,
(S_{E {\bm{k}}})\label{eq:Sigmaperp}. \hspace{3em}
\end{eqnarray}
\end{subequations}
Equation\ (\ref{eq:Sigmaperp}) shows that scattering mixes the
distributions of conserved and precessing spins. This is so because
when one spin at wave vector ${\bm k}$ and precessing about ${\bm
\Omega}_{\bm k}$ is scattered to wave vector ${\bm k}'$ and
precesses about ${\bm \Omega}_{{\bm k}'}$, its conserved component
changes, a process which alters the distributions of conserved and
precessing spin. Equations (\ref{subeq:S}) can be solved
straightforwardly if one assumes the impurity potential to be
short-ranged, obtaining \cite{dim07ss}
\begin{subequations}
\begin{eqnarray}
& & S_{E {\bm{k}}\|} = \Sigma_{E {\bm{k}}\|} \, \tau_p + P_\| \, (1
- \bar{P}_\|)^{-1} \bar{\Sigma}_{E {\bm{k}}\|} \, \tau_p,
\label{eq:Sparallel} \\ [1ex] & & S_{E {\bm{k}}\perp} =
\frac{\bm{\Omega}_{\bm k} \times (\bm{\Sigma}_{E {\bm{k}}\perp}
\tau_p + P_\perp\, \bar{\bm S}_{E {\bm{k}}\|}) \cdot \bm{\sigma}\,
\tau_p}{2\hbar(1 + \Omega_{\bm k}^2 \tau_p^2/\hbar^2)} -
\frac{(\Sigma_{E{\bm k}\perp} \tau_p + P_\perp\, \bar{S}_{E
{\bm{k}}\|})}{1 + \Omega_{\bm k}^2 \tau_p^2/\hbar^2}
\label{eq:Sperp} .
\end{eqnarray}
\end{subequations}
The correction $S_{E {\bm{k}}\|}$ does not give rise to a spin
current. Inspection of Eq.\ (\ref{eq:Sparallel}) shows that
integrals of the form $\int d\theta \, \hat{\mathcal{J}}^\sigma_{i}
\, S_{E {\bm{k}}\|}$ contain an odd number of powers of ${\bm k}$
and are therefore zero. It can, however, give rise to a
nonequilibrium spin density, since integrals of the form $ \int
d\theta \, \hat{s}^\sigma \, S_{E {\bm{k}}\|}$ contain an even
number of powers of ${\bm k}$ and may be nonzero. Similarly $S_{E
{\bm{k}}\perp}$ does not lead to a nonequilibrium spin density. The
expectation value of the spin operator yields integrals of the form
$\int d\theta \, \hat{s}^\sigma \, S_{E {\bm{k}}\perp}^{(0)}$, which
involve odd numbers of powers of ${\bm k}$ and are therefore zero.
This term does, however, give rise to nonzero spin currents, since
integrals if the form $\int d\theta \, \hat{\mathcal{J}}^\sigma_{i}
\, S_{E {\bm{k}}\perp}$ contain an even numbers of powers of ${\bm
k}$ and may be nonzero. Therefore, in the absence of spin-orbit
coupling in the scattering potential, nonequilibrium spin currents
arise from spin precession (as outlined by Sinova \textit{et~al.}
\cite{sin04}), and nonequilibrium spin densities from the absence of
spin precession. The dominant contribution to the nonequilibrium
spin density in an electric field exists because in the course of
spin precession a component of each individual spin is preserved.
For an electron with wave vector $\bm k$, this spin component is
parallel to ${\bm \Omega}_{\bm k}$. In equilibrium the average of
these conserved components is zero. When an electric field is
applied the Fermi surface is shifted and the average of the
conserved spin components may be nonzero, as illustrated in Fig.~1.
This argument explains why the nonequilibrium spin density $\propto
\tau_p^{-1}$ and \emph{requires} scattering to balance the drift of
the Fermi surface. Although spin densities in electric fields
require band structure spin-orbit interactions and therefore spin
precession, the dominant contribution arises as a result of the
absence of spin precession.

\begin{figure}[tbp]
 \includegraphics[width=0.95\columnwidth]{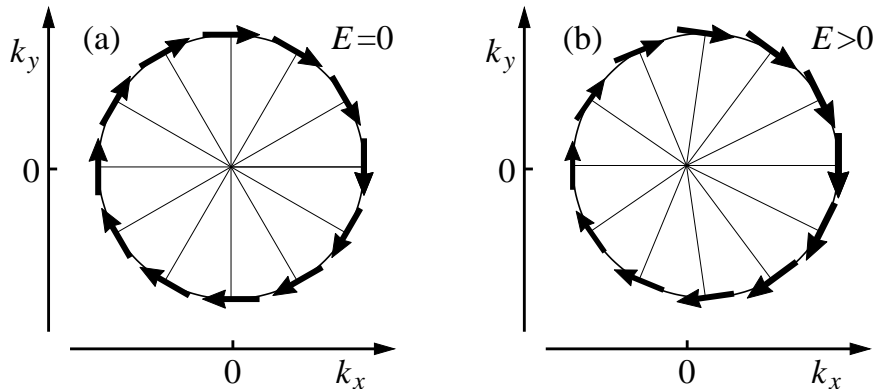}
 \hfill
 \caption{\label{fig:sia_e} Effective field ${\bm \Omega}_{\bm k}$
 at the Fermi energy in the Rashba model \cite{ras84} (a) without ($E=0$) and (b) with an external electric field ($E>0$).}
\end{figure}

Systems in which ${\bm \Omega}_{\bm k}$ is linear in ${\bm k}$ are
special, in that the spin current as defined in this section
vanishes \cite{ino04, dim05, sch02, mis04, kha06, shy06, ll06,
mal05}. This is because of the renormalization of the driving term
on the RHS of Eq.\ (\ref{eq:Sigmaperp}) for $S_{E{\bm k}\perp}$, in
other words because of scattering from the conserved spin
distribution to the precessing spin distribution. In Eq.
(\ref{eq:Sperp}) it is also clear that if $\Sigma_{E{\bm k}\perp}
\,\tau_p + P_\perp\, \bar{S}_{E {\bm{k}}\|}$ vanishes, then all the
contributions to $S_{E {\bm{k}}\perp}$ also vanish. Since
$\bar{S}_{E {\bm{k}}\|}$ effectively represents a steady-state spin
density, we see that the presence of this spin density tends to
diminish the spin current. In systems with energy dispersion linear
in ${\bm k}$ it cancels the spin current completely.

\section{Future Directions}

Whereas the community appears to be in agreement that spin currents
exist and are measurable, many questions remain unanswered.
Theoretically, intrinsic and extrinsic effects (such as due to skew
scattering and side jump) have not been studied on the same footing
for an arbitrary form of band structure spin-orbit interactions. The
relative magnitude of intrinsic and extrinsic spin currents in such
a general system remains to be determined. Also, different
definitions of the spin current give results that often differ by a
sign \cite{shi06, sug06}. The relationship between spin current and
spin accumulation at the boundary is not clear, again thanks to the
non-conservation of spin. It appears that what happens at the
boundary is sensitive to the type of boundary conditions assumed.
Thus so far as quantitative interpretation of experimental data is
concerned, theory has some way to go.

Despite tremendous progress, experiment is still searching for a
reliable way to \emph{measure}, as opposed to \emph{detect}, spin
currents directly. Practically, the question of what to do with spin
once it has been transported/generated remains. The revolutionary
electronic device that harnesses spin currents for a practical
purpose remains to be made, and the challenge of its design
confronts experimentalists and theorists alike.

The research at Argonne National Laboratory was supported by the US
Department of Energy, Office of Science, Office of Basic Energy
Sciences, under Contract No. DE-AC02-06CH11357.

\end{document}